**Title:** Temperature dependent moiré trapping of interlayer excitons in $MoSe_2$-$WSe_2$ heterostructures

**Author Names:** Fateme Mahdikhanysarvejahany[1], Daniel N. Shanks[1], Christine Muccianti[1], Bekele H. Badada[1], Ithwun Idi[1], Adam Alfrey[1], Sean Raglow[1], Michael R. Koehler[2], David G. Mandrus[3-5], Takashi Taniguchi[6], Kenji Watanabe[7], Oliver L.A. Monti[1,8], Hongyi Yu[9], Brian J. LeRoy[1], and John R. Schaibley[1]

**Author Addresses:**
[1]Department of Physics, University of Arizona, Tucson, Arizona 85721, USA
[2]JIAM Diffraction Facility, Joint Institute for Advanced Materials, University of Tennessee, Knoxville, TN 37920
[3]Department of Materials Science and Engineering, University of Tennessee, Knoxville, Tennessee 37996, USA
[4]Materials Science and Technology Division, Oak Ridge National Laboratory, Oak Ridge, Tennessee 37831, USA
[5]Department of Physics and Astronomy, University of Tennessee, Knoxville, Tennessee 37996, USA
[6]International Center for Materials Nanoarchitectonics, National Institute for Materials Science, 1-1 Namiki, Tsukuba 305-0044, Japan
[7]Research Center for Functional Materials, National Institute for Materials Science, 1-1 Namiki, Tsukuba 305-0044, Japan
[8]Department of Chemistry and Biochemistry, University of Arizona, Tucson, Arizona 85721, USA
[9]Guangdong Provincial Key Laboratory of Quantum Metrology and Sensing & School of Physics and Astronomy, Sun Yat-Sen University (Zhuhai Campus), Zhuhai 519082, China

**Corresponding Author:** John Schaibley, johnschaibley@email.arizona.edu

**Abstract:**

$MoSe_2$-$WSe_2$ heterostructures host strongly bound interlayer excitons (IXs) which exhibit bright photoluminescence (PL) when the twist-angle is near 0° or 60°. Over the past several years, there have been numerous reports on the optical response of these heterostructures but no unifying model to understand the dynamics of IXs and their temperature dependence. Here, we perform a comprehensive study of the temperature, excitation power, and time-dependent PL of IXs. We observe a significant decrease in PL intensity above a transition temperature that we attribute to a transition from localized to delocalized IXs. Astoundingly, we find a simple inverse relationship between the IX PL energy and the transition temperature, which exhibits opposite power dependent behaviors for near 0° and 60° samples. We conclude that this temperature dependence is a result of IX-IX exchange interactions, whose effect is suppressed by the moiré potential trapping IXs at low temperature.

**Main Text:**

## Introduction

The IXs of $MoSe_2$-$WSe_2$ heterostructures comprise an electron in the $MoSe_2$ layer bound to a hole in the $WSe_2$ layer (depicted in Fig. 1a). These IXs have been the subject of intense research in recent years [1,2,3,4,5,6,7,8,9,10,11,12, 13,14,15]; however, a fundamental observation of the IX, the strongly temperature dependent PL intensity, has been largely unexplained. It is known that heterostructures with near 0° (R-type) and near 60° (H-type) twist-angle between layers host optically bright IXs due to the alignment of the electron and hole in crystal momentum space[4], where both the spin-singlet and spin-triplet IXs become bright with distinct optical selection rules[16]. Here, the singlet (triplet) corresponds to the IX configuration with opposite (the same) electron and hole spin orientations. Previous reports on IX dynamics have attributed the temperature dependence of the IX to extrinsic disorder potentials[3,16,17,18,19], but many of these studies were performed on non-hBN encapsulated samples with large disorder. The recent reports on trapping IXs at moiré potentials motivate a new exploration of IX dynamics in the context of the moiré trapped IXs[6,7,10,12,20,21].

## Results

In this article, we investigate the physics of IXs in five different hBN encapsulated $MoSe_2$-$WSe_2$ heterostructures (two R-type and three H-type) by performing temperature, excitation power, and time resolved PL measurements. The insets of Figs. 1b-c show the real space and momentum space band alignments of the layers. In R(H)-type heterostructures, the $\pm$K $MoSe_2$ conduction band nearly aligns with the $\pm$K ($\mp$K) $WSe_2$ valence band in momentum space.

Our first observation is that both R- and H-type heterostructures show a >20-fold decrease in PL intensity with increasing temperature from 1.6 K to ~80 K. (Figs. 1b-c). Figure 1d shows an example of the spectrally integrated IX PL intensity as a function of temperature for a representative R- and H-type sample (see Methods for integration procedure). We note that R-type samples show a single dominant singlet IX peak at low temperature and power which broadens with increasing temperature and power to higher energy (Fig. 1b). At powers up to ~500 µW, our H-type samples show two-peaks (for example 1.37 eV and 1.4 eV at 1.6 K in Fig. 1c). We assign the 1.4 eV peak to the triplet IXs as previously reported[16,22]. The nature of the 1.37 eV peak is not completely understood, but its power, polarization dependence and g-factor are similar to that of the 1.4 eV triplet (See Supplementary Fig. 1 and Supplementary Table 1). We therefore refer to the 1.37 eV peak as the triplet', which may originate from IXs localized to lower energy moiré sites[7]. We note that other recent works have observed a similar peak near ~1.35-1.37 eV in H-type samples[22,23,24]. We note that at powers below 500 nW and temperatures below 10 K we observe narrow (~200 μeV) IX emission lines consistent with Seyler et al.[6] , which are inhomogeneously broadened (see Supplementary Fig. 1).

Our second observation is that this transition temperature is excitation power dependent (as shown in Fig. 1e). To elucidate these effects, we performed a systematic investigation of the PL intensity by varying the excitation power for each temperature measured. For each excitation power, we extract a transition temperature, defined as the temperature at which the spectrally integrated IX PL counts are reduced by a half. We emphasize that while previous works have explored the IX temperature[176,17,23,23]dependence, our systematic approach on five different samples allows us to observe a previously obscured and striking feature: the IX transition temperature is inversely related to the IX energy (Fig. 1e).

Figure 1e shows the power dependent transition temperature for five different samples. Because the IX center energy varies with temperature and excitation power, we plot the transition temperature for each power used (~0.01 to 144 μW), and the solid dot marks the transition temperature for the highest excitation power. Figure 1e shows two important features: 1) the transition temperature increases with decreasing IX energy, and 2) the R- and H-type samples show the opposite power dependence. Further analysis of the IX center energy and linewidth as a function of power and temperature is shown in Supplementary Figure 2 and 3.

The energy dependent transition temperature can be explained by the trapping of IXs to moiré potentials below a transition temperature ($T_c$), depicted in Fig. 1f. IXs are localized to moiré traps[21] at $T < T_c$. At the transition temperature ($T=T_c$), thermal energy is sufficient to scatter the IX out of the moiré traps, but the exact value of $T_c$ is dependent on the IX energy of the sample. That is the lower the IX energy, the more thermal energy is required to free the IX from the moiré trap. For $T > T_c$, the IXs are free of the moiré traps, and the strong dipole-dipole interaction between IXs results in the density-dependent expansion of the IX cloud[4], scattering the IX out of the collection spot where the density is high. This results in an effective nonradiative decay of the collected IX PL and a decrease in the integrated PL intensity and lifetime. We note that previous studies in non-encapsulated $MoSe_2$-$WSe_2$ heterostructures[4,17] typically exhibit (low power) linewidths on the order of 50 meV, and different samples exhibit ~80 meV of variation in the IX center energy, which can be attributed to disorder arising from the interactions with the substrate. More recently hBN encapsulated $MoSe_2$-$WSe_2$ heterostructures have been reported which exhibit significantly narrower IX linewidths on the order of 5-10 meV[25,26,27], which were attributed to the ultra-flat hBN substrate and cleaner TMD heterostructures. Here, we investigate hBN encapsulated samples, allowing us to measure the intrinsic IX energy more accurately. We also note that the energies of the two different structure types (R-type vs. H-type) are consistent with the theoretical prediction that R-type heterostructures have a deeper moiré trapping potential than H-type heterostructures[7,10]. We attribute the variation between the different structures of the same type (i.e., the two different R-type samples and three different H-type samples) to variations in doping and small built in electric fields that shift the IX energy [4,9,26,28]. Supplementary Figure 4 shows spatially resolved PL spectra which has relatively small ~1 meV variation in IX center energy.

Figure 1e also shows that the R-type structures exhibit a decrease in transition temperature for increasing excitation power, whereas the H-type structures exhibit an increase in transition temperature for increasing power. To elucidate this effect, we consider the power and temperature dependent IX PL maps for example R- and H-type heterostructures, as shown in the insets of Figs. 2a-b. In both cases, the PL is strongest at low temperature and high power; however, Figs. 2a-b show that the transition temperature decreases for the R-type and increases for the H-type as the excitation power is increased. For the H-type samples which show two PL peaks, we fit each spectrum with two Gaussians to calculate the two peak areas independently. Figure 2b shows the transition temperature for the low (~1.355 eV) peak of the sample $H_3$. The transition temperature of the high energy 1.39 eV peak also increases with increasing power and is shown in Supplementary Fig. 5. The measurements were repeated on all five samples, whose data are shown in Supplementary Fig. 5 and are consistent with the qualitative behaviors shown in Figs. 2a-b.

In order to explore this qualitatively different power dependence between R- and H- type samples, we performed time resolved PL spectroscopy on the samples (see Supplementary Fig. 6). Figures 2c-d show the temperature and lifetime dependence of PL. The time-dependent PL is fit with a bi-exponential for each temperature whose time constants ($T_1$ and $T_2$) are plotted in the insets of Figs. 2c-d. The R(H)-type shows a short lifetime component $T_1$ on the order of 200 ns (12 ns) at low temperature, which decreases near the high power transition temperatures (40 K and 33 K, respectively). We attribute this fast decay component to the scattering time of IXs out of the moiré potential. The longer lifetime for the R-type structures is consistent with the lower IX energy and deeper confinement potential. Both structures exhibit a long lifetime component ($T_2$) on the order of 300-800 ns that does not vary significantly with temperature (second insets of Figs. 2c-d). The long lifetime component is attributed to the intrinsic radiative and non-radiative decay. We emphasize the primary finding is that at high temperature, the lifetime decreases while the integrated IX PL intensity also decreases, which is due to the activation of a new nonradiative decay channel, i.e., the scattering of the IX outside of the collection spot when they are no longer localized to moiré potentials.

In order to understand the opposite power dependent transition behaviors for R- and H-type structures, we investigated the low temperature (1.6 K) power dependence of the IX (Fig. 3). The IXs in our experiment are generated through the laser excitation of the intralayer excitons followed by the ultrafast interlayer charge transfer. It has been shown that the interlayer charge transfer largely conserves the spin of the electron/hole, thus the formed IXs in both R- and H-type structures should be dominated by the intravalley or intervalley spin-singlet (See Supplementary Note 1). In the R-type structure, depicted in Fig. 3, the dominant IX species is the lowest energy intravalley singlet, composed of an electron and hole of opposite spin in the same valley. Figure 3b shows (normalized) low- and high-power PL spectra of an R-type structure at low temperature (1.6 K). At 400 nW excitation, a Lorentzian fit to the PL spectrum yields a FWHM of ~8 meV, which is consistent with other recent reports[25,27,26,28]. At high power, the IX lineshape broadens asymmetrically to higher energy, which is shown in Figs. 3c-d for two different R-type samples. We attribute this broadening to the filling of higher energy IX states that are localized to the moiré traps. This effect is more prominent for the R-type due the deeper moiré potential[7], which is again consistent with the observed lower IX energy for R-type structures which can also host higher energy IX states that are filled at higher IX density.

Figure 3e depicts the band diagram for the H-type structures, where the lowest energy direct transition is the intravalley triplet, composed of an electron and hole of the same spin in the same valley (yellow transition in Fig. 3e). Since the formed IXs should be mostly in the singlet configuration, IXs in H-type structures should be dominated by the lowest energy intervalley singlet, composed of an electron and hole with opposite spin in opposite valleys (depicted as the red transition in Figures 3a,e). Figure 3f shows the (normalized) low- and high-power PL spectra of an H-type structure, where we attribute the ~1.37 eV and ~1.39 eV peaks to the triplet' and triplet IX. We note that the H-type structures also exhibit a higher energy intravalley singlet[16,26,28], composed of an electron and hole in the same valley with the same spin. At very high powers (~ 1 mW), we observe the emergence of the intravalley singlet peak near 1.425 eV (see Supplementary Fig. 1). The intravalley singlet is not observed in our low power measurements and is not required to explain our primary results.

Finally, we consider a model that explains the different power dependent transition behaviors for R- and H-type structures (i.e., the insets of Figs. 2a-b). This difference can be understood by considering the effect of the scattering from dark intervalley IXs to bright intravalley IXs (see Fig. 4). In both structures, optical injection leads to the preferential formation of singlet IXs (see Supplementary Note 1). For R-type samples, the low energy singlet is a bright intravalley singlet as depicted in Figure 3a, whereas for H-type samples, the low energy singlet is a dark intervalley singlet as depicted in Figure 3e.

In Supplementary Note 2, we present a model that reproduces our experimental observations for both R- and H-type samples. The model is based on the IX-IX exchange interaction between pairs of IXs at neighboring moiré sites (see Supplementary Figs. 8-9). In both cases, the scattering is important particularly at higher excitation power and higher temperatures where the IX wavefunction is more delocalized and IX-IX exchange interactions can mediate the scattering process between intervalley IXs and intravalley IXs[4]. In both cases, the IX-IX exchange interaction can scatter singlets to triplets, but since singlets are bright in R-type and dark in H-type samples, this exchange scattering has opposite effects on the PL intensity. In R-type samples, a pair of bright intravalley singlet IXs can scatter to a pair of dark intervalley triplet IXs, which then serves as an additional PL decay channel at high power or high temperature (see Fig. 4a).

On the other hand, for H-type (Fig. 4b), the lowest energy intervalley singlet is a dark state. Scattering from the dark intervalley singlet to the bright intravalley triplet increases the PL intensity. Therefore, the increase of the transition temperature for increasing excitation power (Fig. 2b) is unique to H-type structures and is a consequence of the high density of dark intervalley singlets. At high temperature and excitation power, the dark intervalley singlet can serve as a reservoir of IXs that then scatter into bright intravalley triplets. Supplementary Figure 10 shows theoretical PL intensity as a function of temperature for R- and H-type samples whose power dependence matches our experimental result.

## Discussion

In summary, we have systematically investigated the effect stacking type on IXs in $MoSe_2$-$WSe_2$ heterostructures. We show that the IX energy is inversely related to a temperature dependent transition of the IX PL intensity for both R- and H-type structures. We attribute this transition to the trapping of IXs to moiré potentials at low temperature, which also results in an increase of the IX lifetime. Finally, we explain the importance of the IX-IX exchange interaction on the power dependence of the transition. Our work has important consequences for future studies of IXs in TMD heterostructures because it explains the intrinsic temperature dependence of IXs, which was not previously understood. We also show that R- and H-type structures have opposite behaviors around this transition that make the identification of heterostructure type important for the interpretation of past and future works.

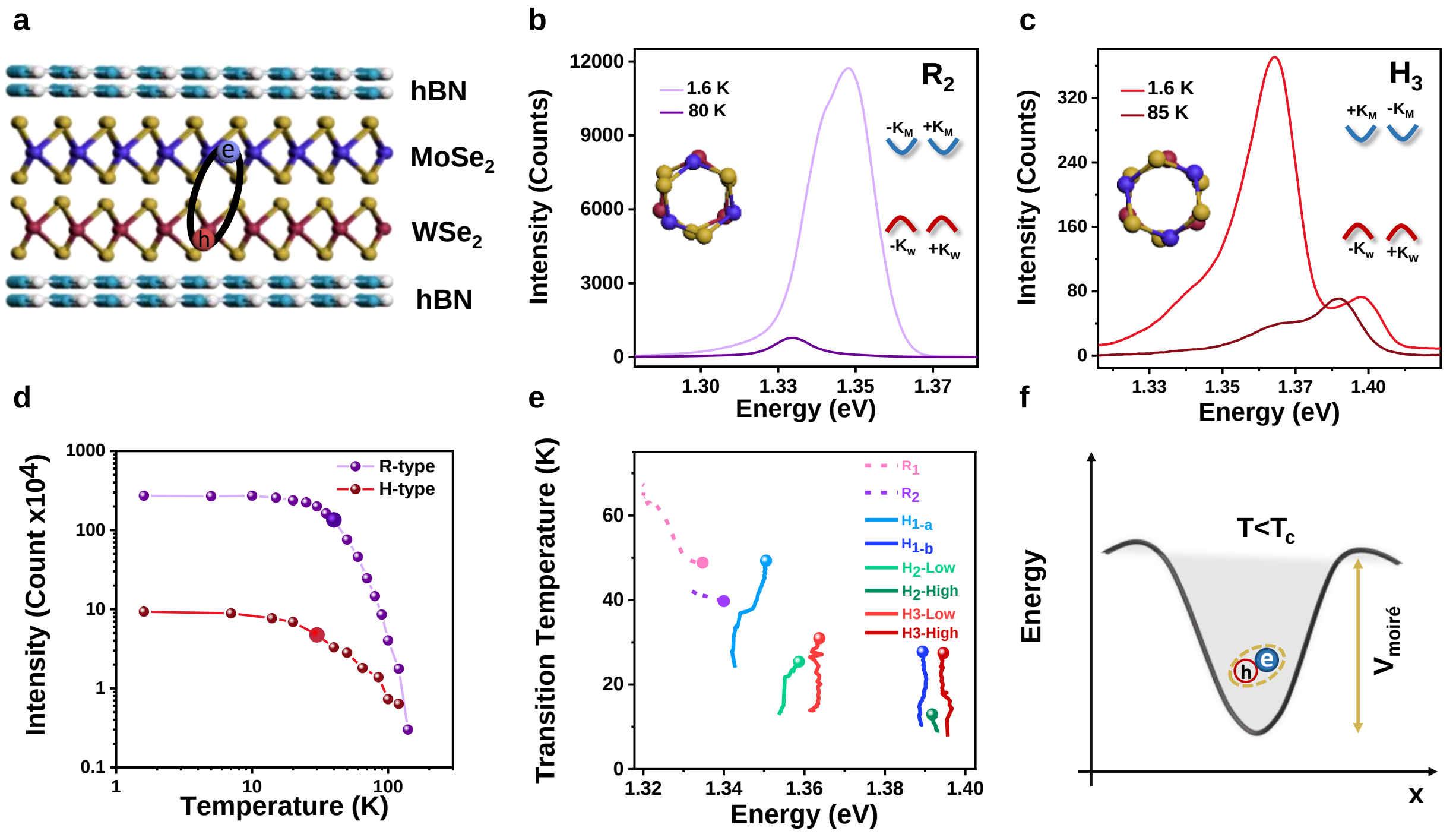

**Fig. 1 |Temperature dependent PL transition from R- and H-type structures. a,** Cartoon depiction of hBN-encapsulated $MoSe_2$-$WSe_2$ heterostructure showing an IX composed of an electron and hole in opposite layers. **b-c,** Example PL spectra of R-type (H-type) IX at two different temperatures for fixed excitation power 56 μW (74 μW). The left insets depict a top view of the real space alignment. The right insets show the alignment of the valleys in momentum space. **d,** Spectrally integrated PL of R- and H-type heterostructures as a function of temperature, showing temperature dependent transitions at 40 K and 30 K, respectively, marked by larger circles. **e,** Energy dependent transition temperature for five different samples. The R and H subscripts label the sample number, labelled in order of increasing IX energy, where the a and b subscripts denote two disconnected sample regions of the same device. The IX center energy varies with temperature and excitation power, the transition temperature is plotted for each power, and the solid dot marks the transition temperature for the highest excitation power. **f,** Cartoon of IX trapped in moiré potential ($V_{moiré}$) below a transition temperature ($T_c$). Excitation wavelength is 670 nm, and the confocal collection diameter is 1 μm.

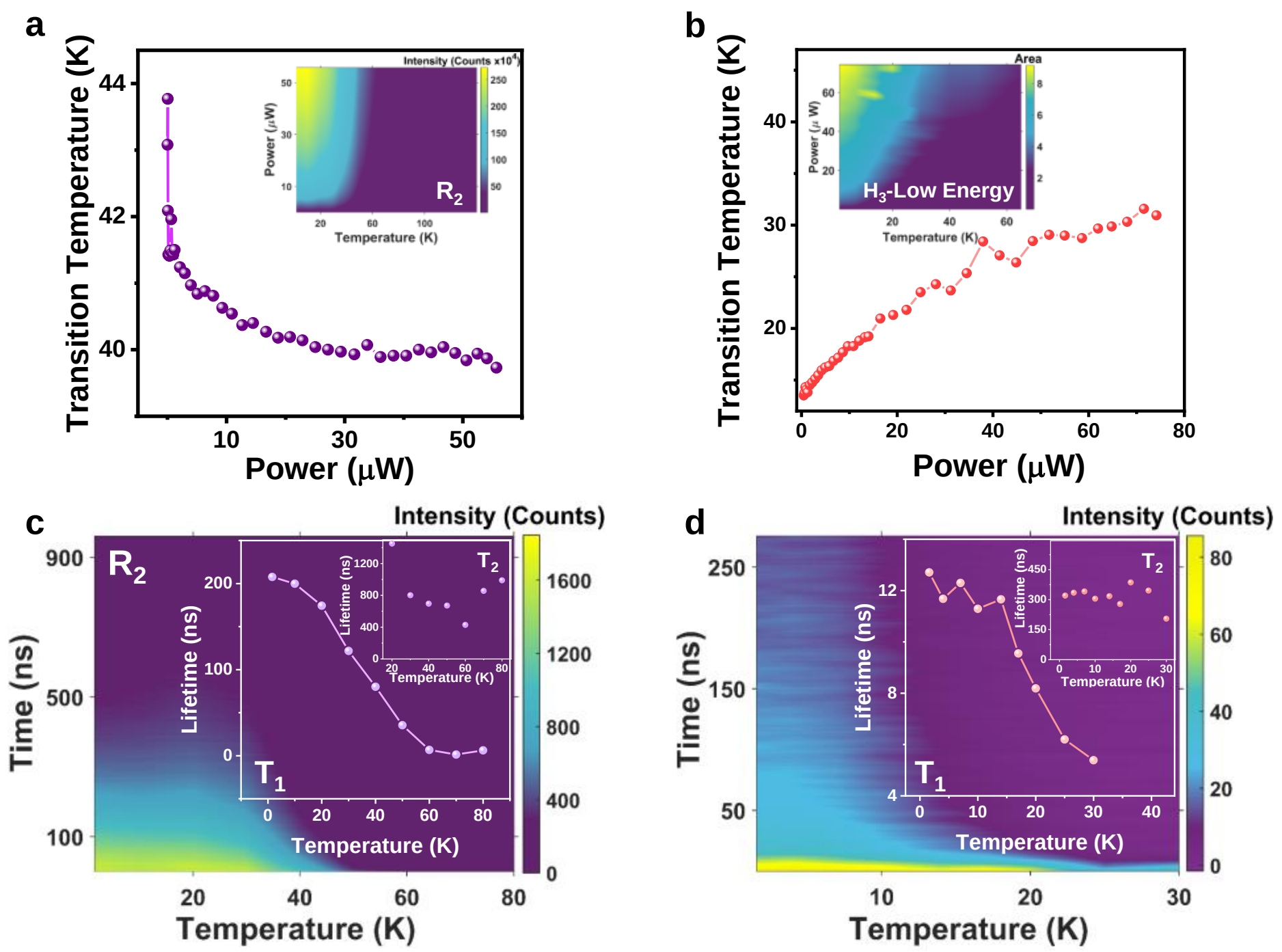

**Fig. 2 |Twist-angle dependent transition temperature for R- and H-type heterostructures. a-b,** Transition temperature as a function of excitation power for an R-(H-)type sample, which decreases (increases) with increasing power. The insets show the full temperature and power dependent integrated IX PL maps. **c-d,** Temperature and lifetime dependent integrated IX PL map for an R-(H-)type sample. The insets show the results of a biexponential fit for each temperature. The short lifetime $T_1$ is on order of 200 ns (12 ns) for R- (H-type) samples at low temperature. In both cases, the lifetime decreases with increasing temperature. The long lifetime $T_2$ component is on order of 900 ns (300 ns) and is relatively constant with temperature.

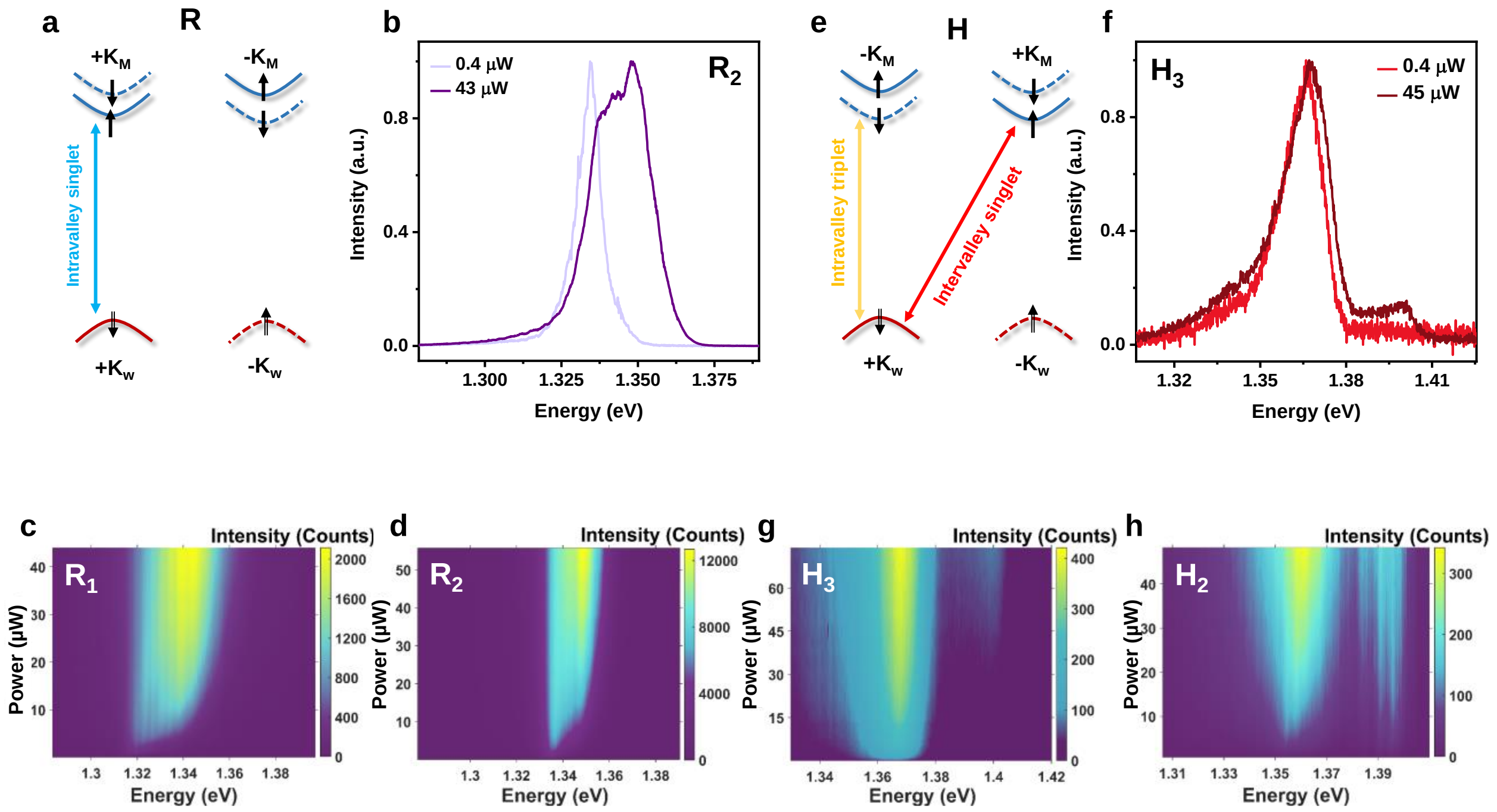

**Fig. 3 |Energy diagrams and power dependence of R- and H-type heterostructures. a,** Energy diagram of IXs in R-type heterostructures showing the ±K valley conduction band of the $MoSe_2$ nearly aligning with the ±K valley valence band of the $WSe_2$. The solid arrow depicts the spin of the electrons, the split arrow depicts the spin of the holes, and the K subscripts label the layer. One of the optically bright intravalley singlet transitions is shown in blue between the +K valley of both $MoSe_2$ and $WSe_2$ (i.e., $+K_W$ to $+K_M$). **b,** Example of normalized R-type PL spectra at low and high excitation powers. **c-d,** Power dependent PL spectra for two different R-type samples show the onset of a high energy shoulder with increasing power. **e,** Energy diagram of IXs in H-type heterostructures showing the ±K valley conduction band of the $MoSe_2$ nearly aligning with the ∓K valley valence band of the $WSe_2$. The lowest energy direct transition is an intravalley triplet (yellow), composed of an electron and hole of the same spin in the same valley. The red arrow depicts the intervalley singlet transition. **f,** Example of normalized H-type PL spectra at low and high excitation powers. **g-h,** Power dependent PL spectra for two different H-type samples showing the emergence of a spectrally isolated peak at the intravalley singlet energy at high power highlighted by the dashed white lines. T=1.6 K for all measurements.

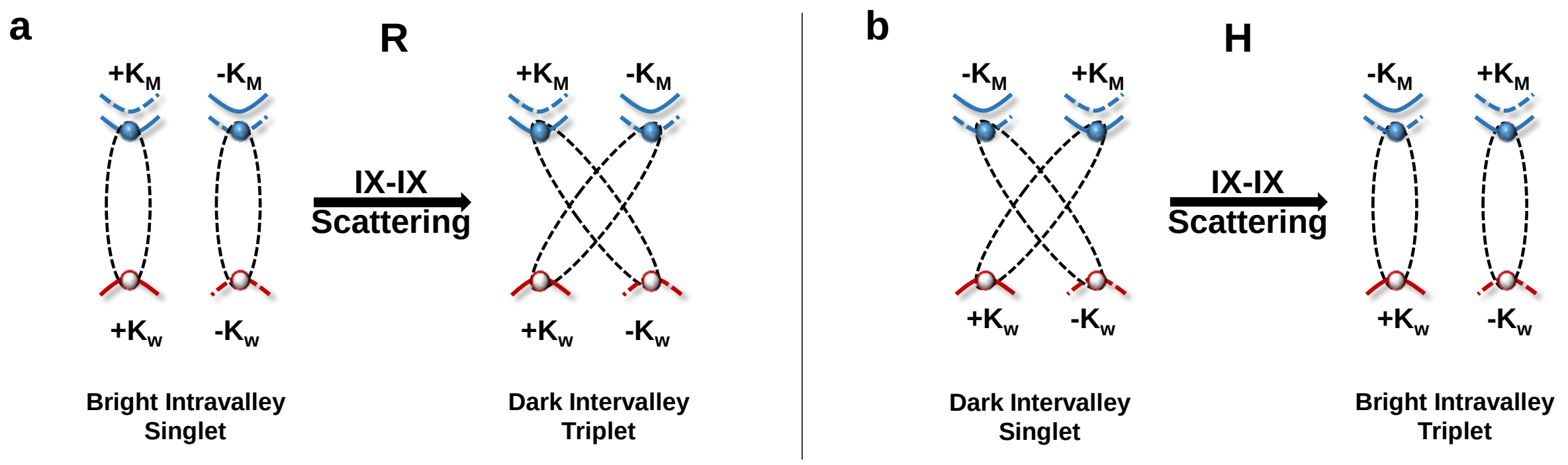

**Fig. 4 | IX-IX exchange interaction for R- and H-type heterostructures. a,** Depiction of R-type decay channel where the IX-IX exchange interaction can scatter bright intravalley singlet IXs to dark intervalley triplet IXs. **b,** Depiction of H-type decay channel where the IX-IX exchange interaction can scatter dark intervalley singlet IXs to bright intravalley triplet IXs.

## Methods:

### Sample fabrication:

$MoSe_2$, $WSe_2$, and hBN flakes were exfoliated from bulk crystals onto Si/$SiO_2$ wafers using Scotch tape. Layer thickness was measured by optical contrast, atomic force microscopy, and PL. The heterostructures were assembled using the polymer based dry transfer method[29]. We used thick (> 20 nm) bottom hBN to ensure the highest heterostructure quality. Polarization-resolved second harmonic generation was[30,31] used to align monolayers of $MoSe_2$ and $WSe_2$ near 0° or 60°. The R- and H-type heterostructures were distinguished by performing low temperature and low power Zeeman splitting measurements on the moiré excitons following the work of Seyler et al. (Supplementary Table. 1 and Supplementary Fig. 1)[6].

### Optical experiments:

Temperature dependent PL measurements were performed in an optical cryostat by exciting the sample with a 670 nm vertically polarized continuous wave (CW) diode laser focused to a diffraction limited spot with a 0.81 NA objective. We used a confocal pinhole to specifically collect IX emission from the ~ 1 μm excitation spot. The PL spectra were detected with a grating spectrometer and cooled CCD camera.

To analyze the transition temperature, we spectrally integrate the PL in two different ways. In the samples with only dominant one peak, we integrated over a range of ~100 meV centered on the peak in order to define the transition temperature. Whereas for the H-type samples, that exhibit triplet' and triplet peaks, which have a resonance near 1.36 eV and 1.39 eV at low temperature, we fit two Gaussians to each spectrum and used the area for each peak, determined by the fit, to calculate the transition temperature. For each power, the transition temperature was calculated as the temperature when the integrated area was half of the value at 1.6 K. The center energies plotted in Figure 1e were then determined by fitting to spectrum at the transition temperature for each power. The temperature dependent data were recorded from a single thermal cycle. However, the measurements for several samples were repeated for different thermal cycles and produced the same results.

The time-resolved PL measurements were performed using a time correlated single photon counting technique. The samples were excited with 20-30 ns square pulses generated by optically gating a 720 nm CW laser with an acousto-optic modulator. The PL was spectrally filtered using a single grating spectrometer and detected with a single photon avalanche detector and Picoharp picosecond event timer (see Supplementary Fig. 6).

**Data availability:**

The data that support the findings of this study are available from the corresponding author upon reasonable request.

**Code availability:**

Upon request, authors will make available any previously unreported computer code or algorithm used to generate results that are reported in the paper and central to its main claims.

**Acknowledgments:**

**General:** We acknowledge useful discussions with Rolf Binder and Allan MacDonald.

**Funding:** This work is mainly supported by the National Science Foundation (Grant Nos. DMR-1838378 and DMR- 2054572). DGM acknowledges support from the Gordon and Betty Moore Foundation's EPiQS Initiative, Grant GBMF9069. K.W. and T.T. acknowledge support from the Elemental Strategy Initiative conducted by the MEXT, Japan, Grant No. JPMXP0112101001, JSPS KAKENHI Grant No. JP20H00354 and the CREST(JPMJCR15F3), JST. JRS acknowledges support from AFOSR (Grant Nos. FA9550-17-1-0215, FA9550-18-1-0390, and FA9550-20-1-0217) and the National Science Foundation Grant. No. ECCS-1708562. BJL acknowledges support from the National Science Foundation under Grant No. ECCS-1607911, and Army Research Office under Grant no. W911NF-18-1-0420. HY acknowledges support by the Department of Science and Technology of Guangdong Province in China (2019QN01X061).

**Author Contributions**:

JRS and BJL conceived and supervised the project. FM fabricated the structures and performed the experiments, assisted by DNS, BHB, MK, CM, II, AA and SR. FM analyzed the data with input from JRS and BJL. MRK and DGM provided and characterized the bulk $MoSe_2$ and $WSe_2$ crystals. TT and KW provided hBN crystals. HY provided theoretical support in interpreting the results. FM, JRS, BJL, and HY wrote the paper with input from OLAM. All authors discussed the results.

**Competing Interests:**

The authors declare no competing interests.

**Supplementary Information for**

# Temperature dependent moiré trapping of interlayer excitons in $MoSe_2$-$WSe_2$ heterostructures

**Author Names:** Fateme Mahdikhanysarvejahany[1], Daniel N. Shanks[1], Christine Muccianti[1], Bekele H. Badada[1], Ithwun Idi[1], Adam Alfrey[1], Sean Raglow[1], Michael R. Koehler[2], David G. Mandrus[3-5], Takashi Taniguchi[6], Kenji Watanabe[7], Oliver L.A. Monti[1,8], Hongyi Yu[9], Brian J. LeRoy[1], and John R. Schaibley[1]

**Author Addresses:**
[1]Department of Physics, University of Arizona, Tucson, Arizona 85721, USA
[2]JIAM Diffraction Facility, Joint Institute for Advanced Materials, University of Tennessee, Knoxville, TN 37920
[3]Department of Materials Science and Engineering, University of Tennessee, Knoxville, Tennessee 37996, USA
[4]Materials Science and Technology Division, Oak Ridge National Laboratory, Oak Ridge, Tennessee 37831, USA
[5]Department of Physics and Astronomy, University of Tennessee, Knoxville, Tennessee 37996, USA
[6]International Center for Materials Nanoarchitectonics, National Institute for Materials Science, 1-1 Namiki, Tsukuba 305-0044, Japan
[7]Research Center for Functional Materials, National Institute for Materials Science, 1-1 Namiki, Tsukuba 305-0044, Japan
[8]Department of Chemistry and Biochemistry, University of Arizona, Tucson, Arizona 85721, USA
[9]Guangdong Provincial Key Laboratory of Quantum Metrology and Sensing & School of Physics and Astronomy, Sun Yat-Sen University (Zhuhai Campus), Zhuhai 519082, China

**Corresponding Author:** John Schaibley, johnschaibley@email.arizona.edu

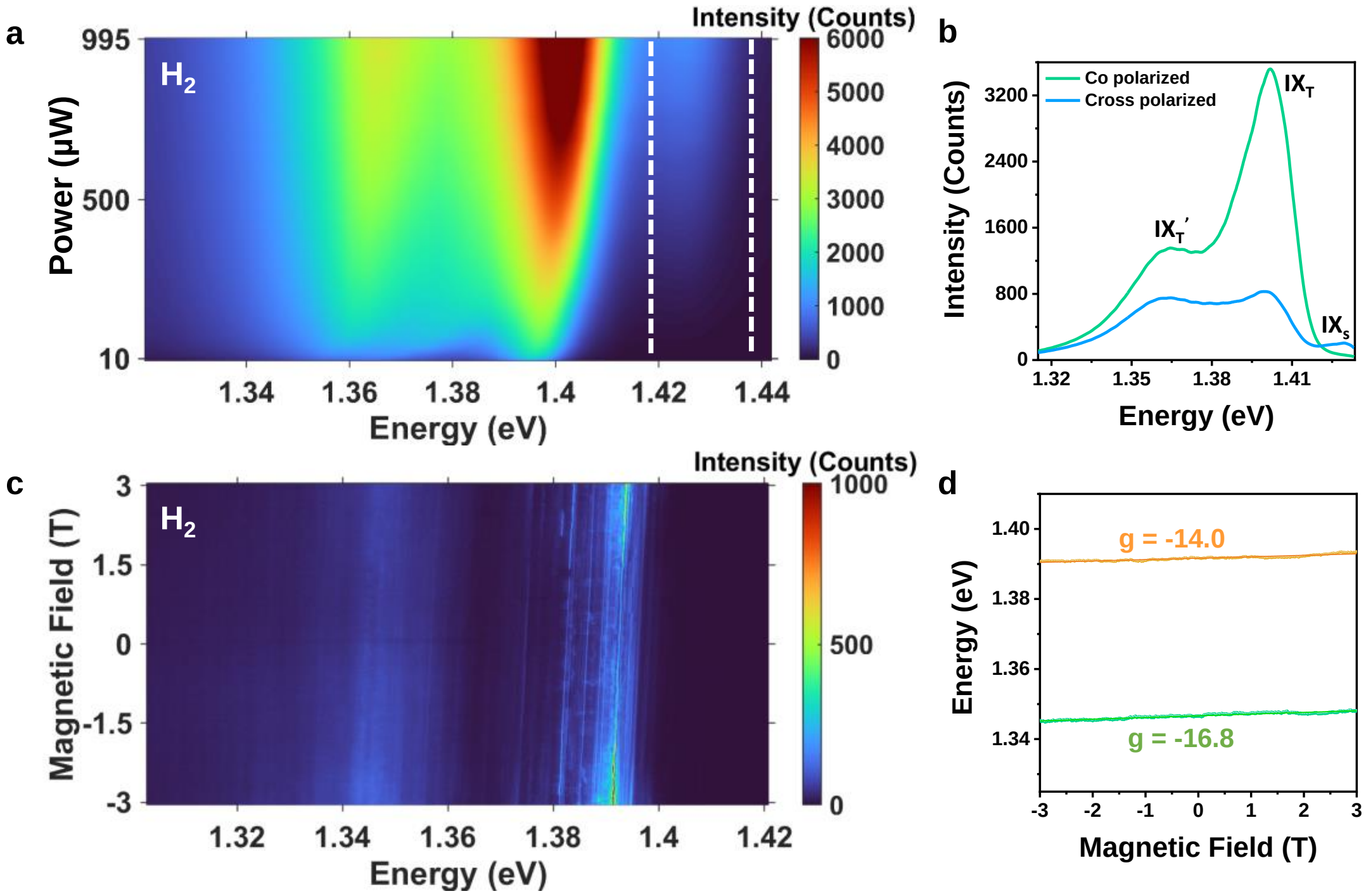

**Supplementary Fig. 1| Extended data from $H_2$ sample. a,** PL as function of excitation power under 1.72 eV excitation (near the $WSe_2$ exciton resonance). We identify the singlet near 1.425 eV at high power. **b,** Co- and Cross- circularly polarized PL when exciting with 1.72 eV. Triplet ($IX_T$ and $IX_T$') peaks are co-polarized with the excitation laser, whereas the singlet ($IX_s$) is cross-polarized. **c,** $\sigma^-$ polarized PL as a function of applied magnetic field at low power (150 nW). **d**, Extracted center energy as function of magnetic field for example ~1.35 eV (green) and ~1.39 eV (orange) peaks. The extracted g-factors are labelled.

| Device | $\Delta\theta$ By edges | $\Delta\theta$ By SHG | g-factor |
|---|---|---|---|
| $R_1$ | 7±(1) | | 8.3 |
| $R_2$ | | 1.3±(0.4) | 8.0 |
| $H_1$ | 4.8±(0.3) | | |
| $H_2$-Low Energy | 4.2±(0.5) | | -16.8 |
| $H_2$-High Energy | 4.2±(0.5) | | -14.0 |
| $H_3$ | | | -16.2 |

**Supplementary Table 1| Estimated twist angles and measured Zeeman shifts.** Three different methods for estimating the exact twist-angles were used. For three devices, we identified long edges of the constituent monolayers by optical microscopy after the heterostructures were fabricated. In sample $R_2$, which had isolated monolayer and heterobilayer regions, we were able to perform polarization resolved second harmonic generation[1,2,3,4] on the different regions to measure the twist angle. In order to distinguish between R- and H-type structures, we measured the g-factor for four different samples in a magnetic optical cryostat. Supplementary Fig. 1 shows example Zeeman data for an H-type sample. Device $H_1$ was destroyed before Zeeman measurements were performed, so we assigned the type based on this sample's energy, power, and temperature dependence.

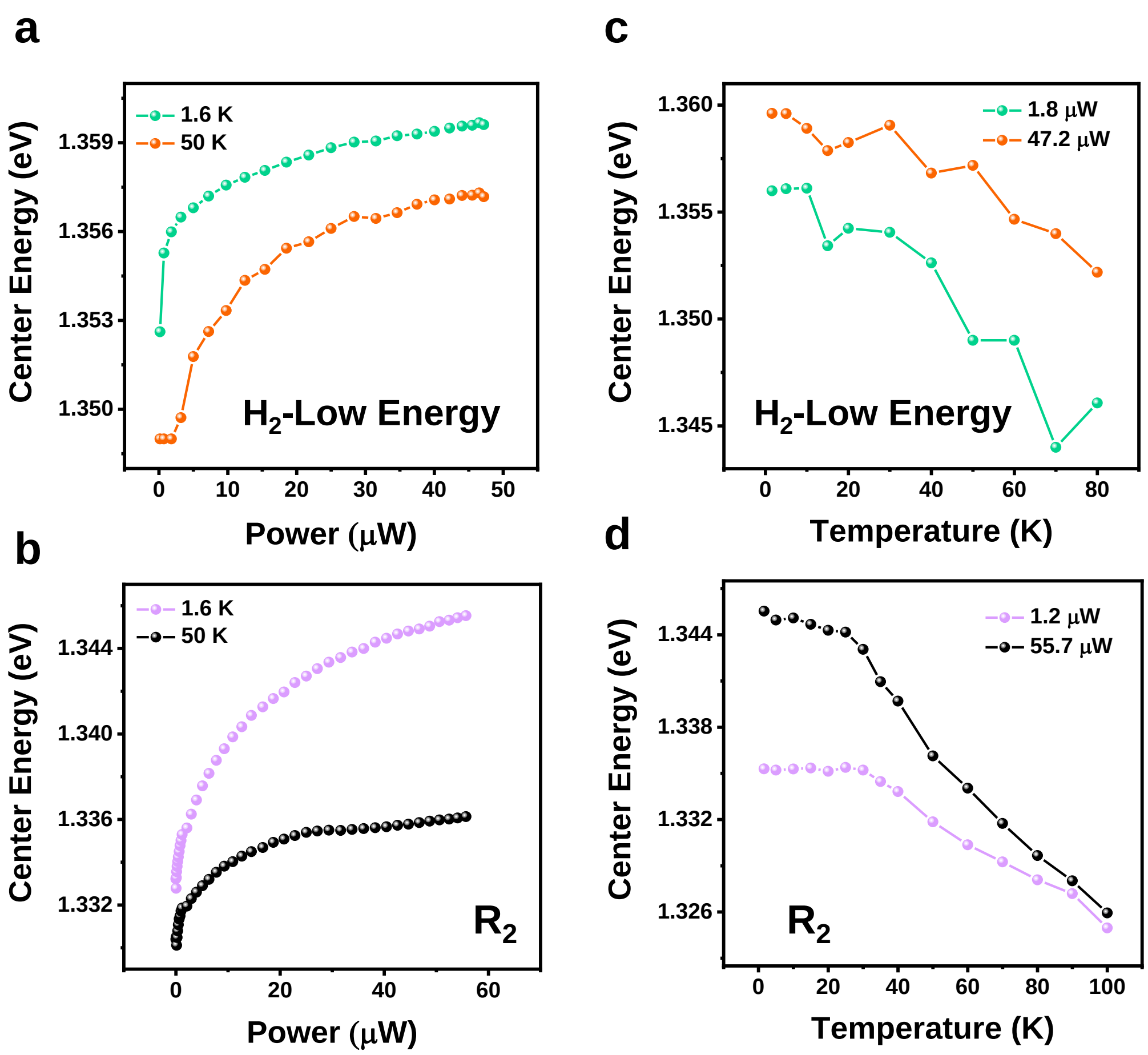

**Supplementary Fig. 2| Center energy as function of temperature and power. a-b,** Center energy as a function of power for $H_2$ and $R_2$ with fixed temperature (1.6 K and 50 K). **c-d,** Center energy as a function of temperature for $H_2$ and $R_2$ with fixed powers (~1 μW and ~50 μW).

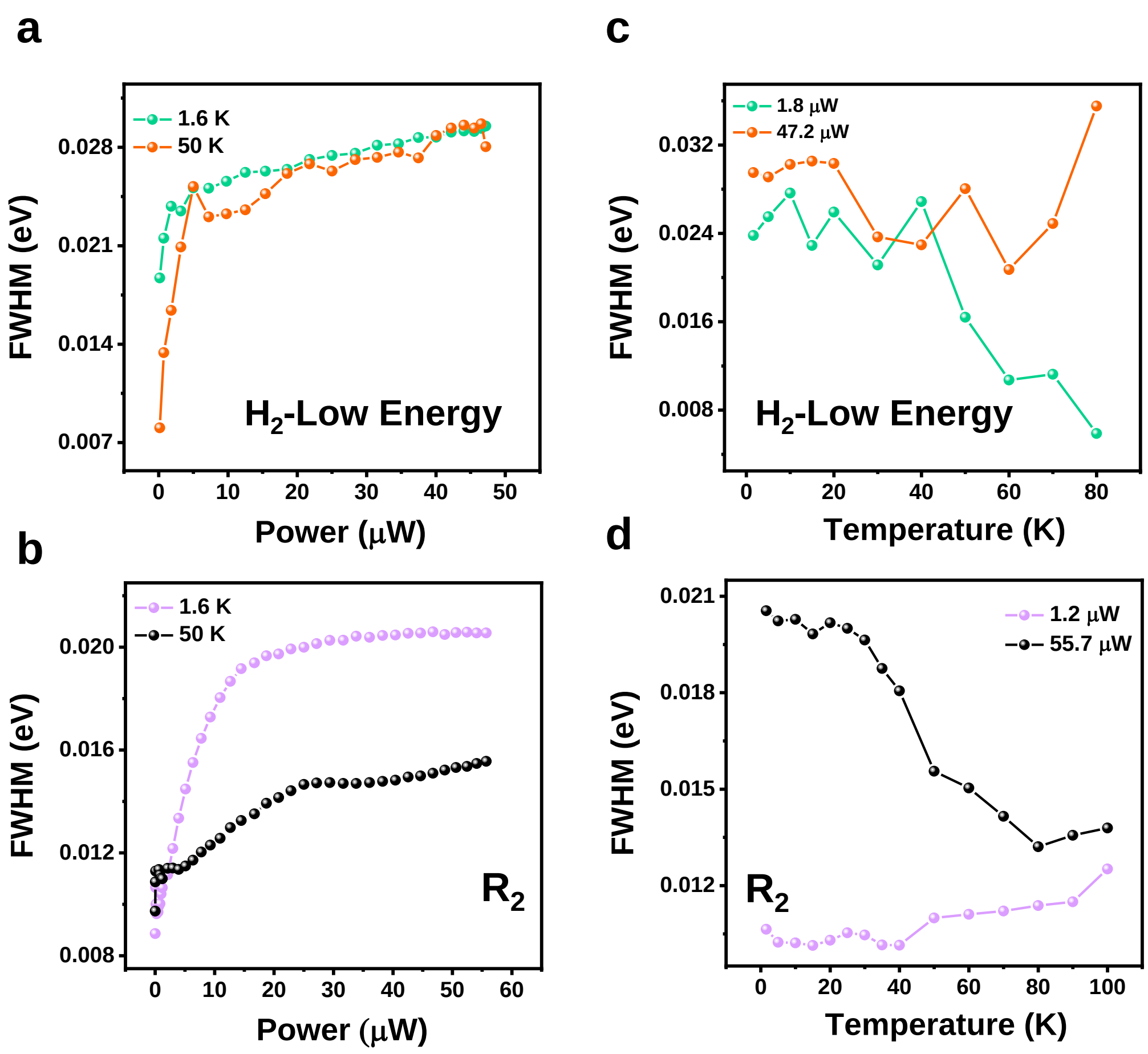

**Supplementary Fig. 3| IX linewidth as a function of temperature and power. a-b,** Linewidth (**full width half max-FWHM**) as a function of power for $H_2$ and $R_2$ with fixed temperature (1.6 K and 50 K). **c-d,** Linewidth (FWHM) as a function of temperature for $H_2$ and $R_2$ with fixed powers (~1 μW and ~50 μW).

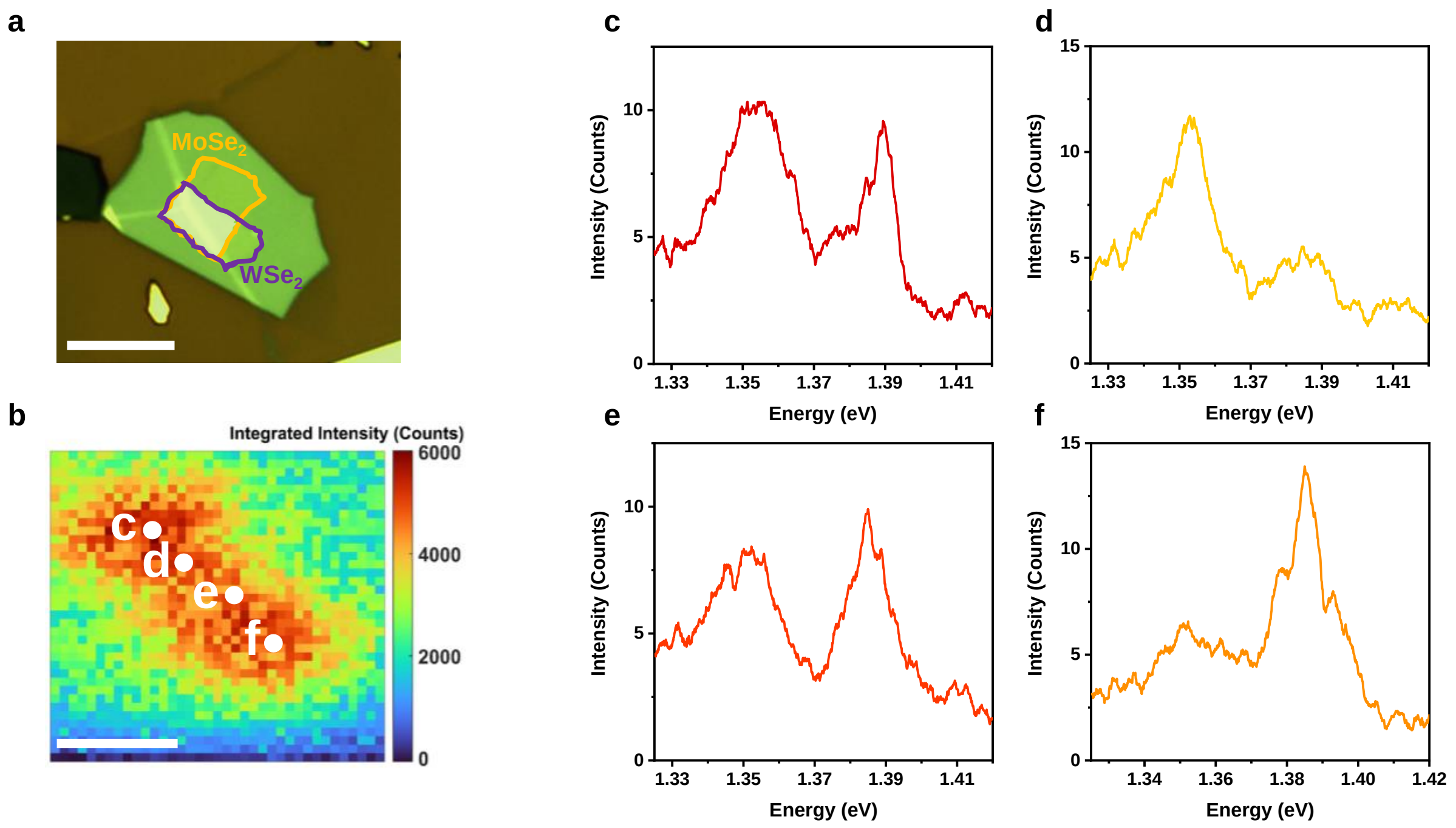

**Supplementary Fig. 4| Spatially resolved PL of sample H2. a,** Optical image of H2 sample with an active area of ~10 μm x 3 μm**.** Scale bar is 5 μm. **b,** Spectrally integrated PL of sample H2 measured with a 1 μm confocal detection diameter. The color bar denotes integrated counts. Scale bar is 2 μm. **c-f,** PL spectrum from the points labeled in **b**. For the spectra shown, the standard deviation of the center energy position is 0.98 meV for the low energy peak, and is 1.9 meV for the higher energy peak.

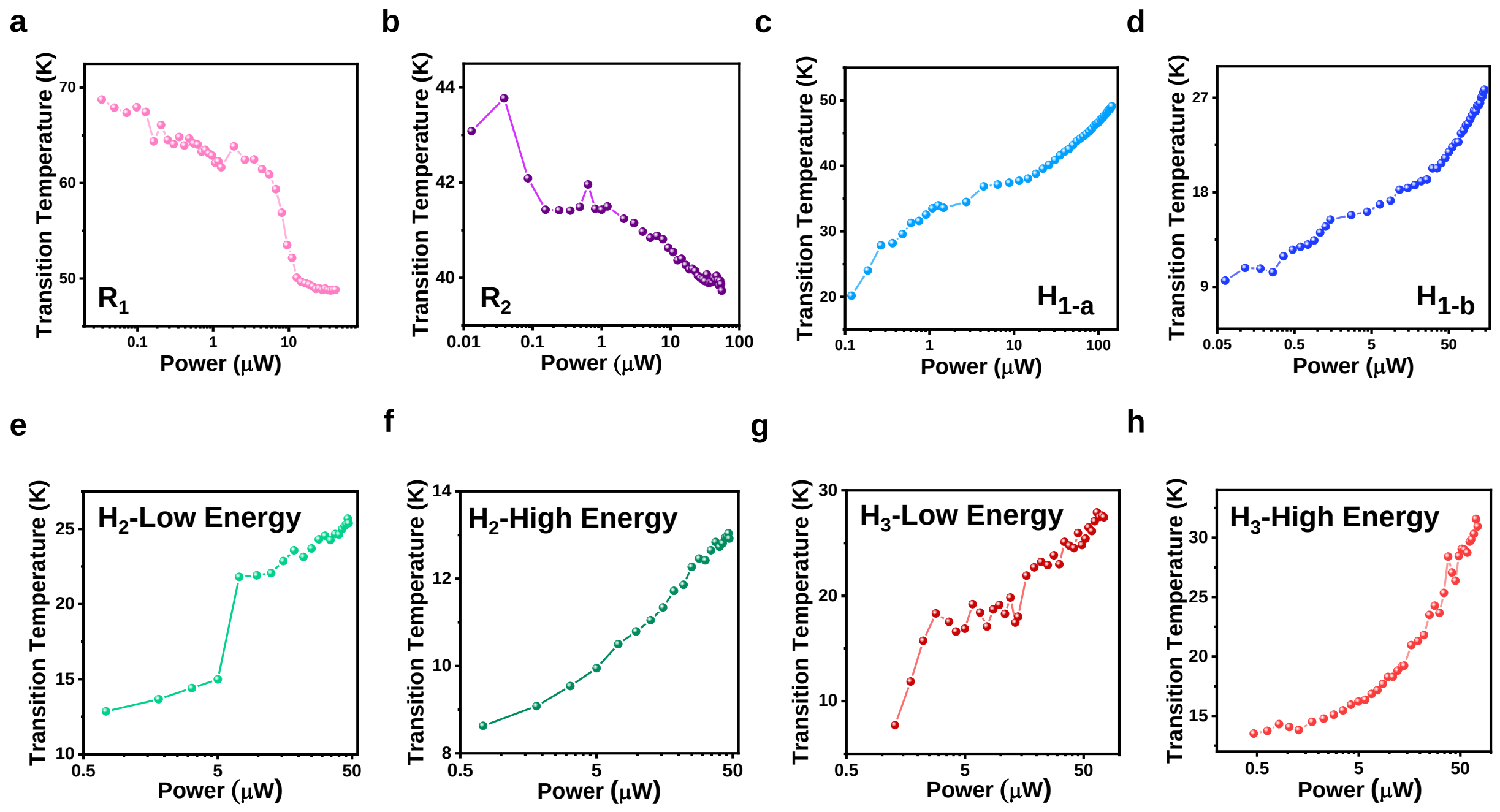

**Supplementary Fig. 5| Power dependent transition temperature on all samples. a-h,** Transition temperature as a function of power plotted on a semi-log scale.

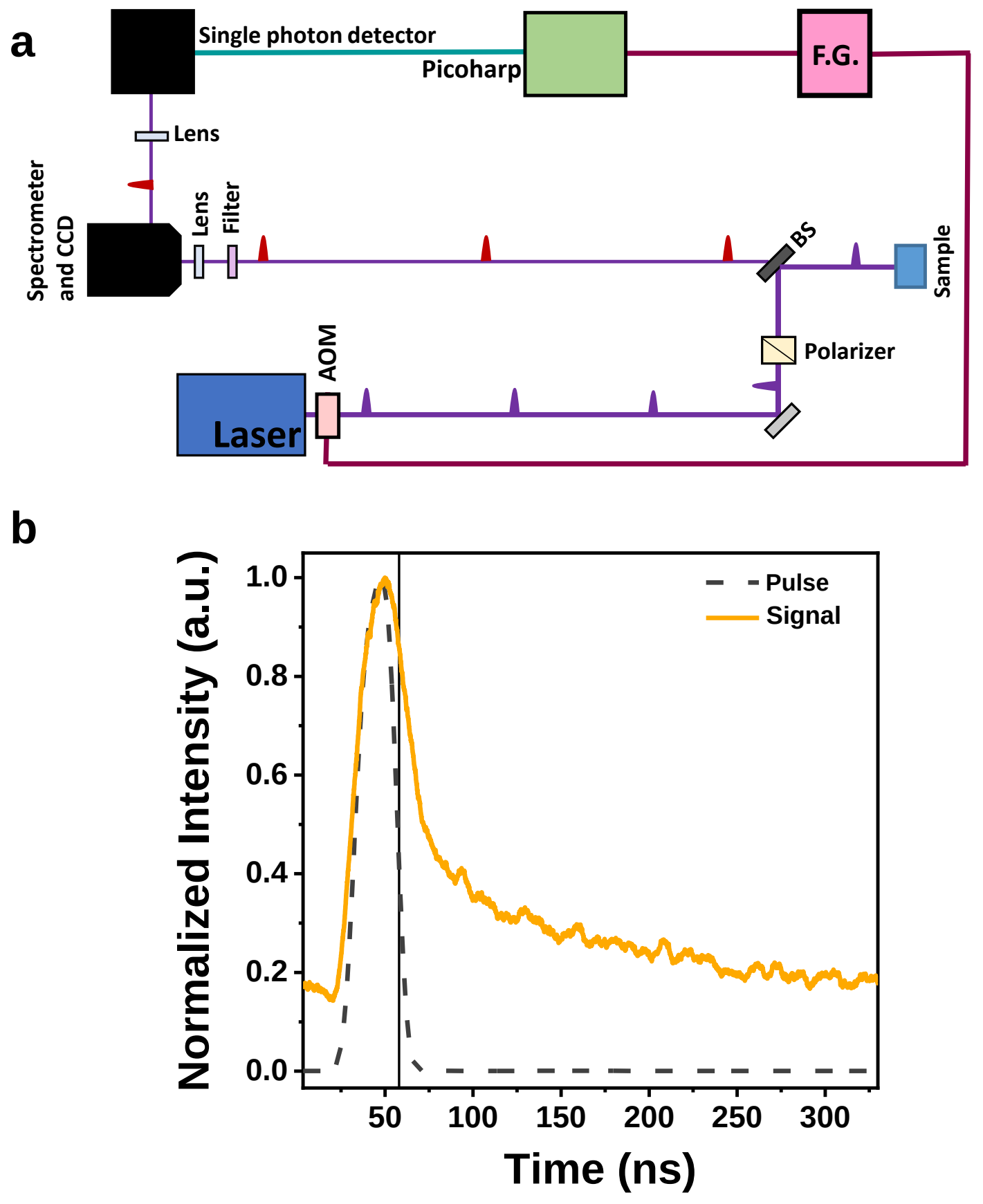

**Supplementary Fig. 6| Schematic of time-resolved experiments and example data. a,** Optical setup used in the time-resolved PL measurements. The system is composed of a CW Ti-sapphire laser modulated by an acousto-optic-modulator (AOM) and a time correlated single photon setup (F.G.- function generator, BS- beam sampler). **b,** Example of the normalized time resolved PL signal of an H-type sample. The dashed line shows the generated square pulse with fast fall time of 2 ns used for exciting the sample. The pulse width is 30 ns and the repetition rate is 300 ns. In order to extract the lifetime, we only analyzed the time resolved PL after the laser intensity dropped below 38% of its peak intensity.

## Supplementary Note 1: IX Singlet Formation

Our analysis is based on the observation that the interlayer charge transfer process largely conserves the electron/hole spin [5,6,7]. As shown in Supplementary Fig. 7, when using a σ+ circularly polarized laser, most of the excited $+K_W$ valley intralayer excitons in $WSe_2$ will relax to the lowest-energy spin-singlet IXs in the ($+K_M$, $+K_W$) configuration. Whereas the lowest energy ($-K_M$, $+K_W$) IX corresponds to a spin-triplet thus is less probable to appear. These ($+K_M$, $+K_W$) IXs are intravalley and optically bright in R-type samples, but intervalley and optically dark in H-type samples.

In our measurement, we used linearly polarized excitations to simultaneously excite both $+K_W$ and $-K_W$ valleys, see Supplementary Fig. 7b. This results in the formation of ($+K_M$, $+K_W$) and ($-K_M$, $-K_W$) IXs, both are spin-singlet and optically dark in H-type samples. The same analysis can be applied to the valley depolarization process. The main valley depolarization mechanism is the electron-hole exchange interaction of the intralayer exciton, which, however, only affects the spin-singlets. So neither the linearly polarized excitation nor the exchange-induced valley depolarization can introduce ($-K_M$, $+K_W$) or ($+K_M$, $-K_W$) spin-triplet IXs to the heterobilayer, and a reservoir of dark intervalley singlet IXs will always appear in H-type samples.

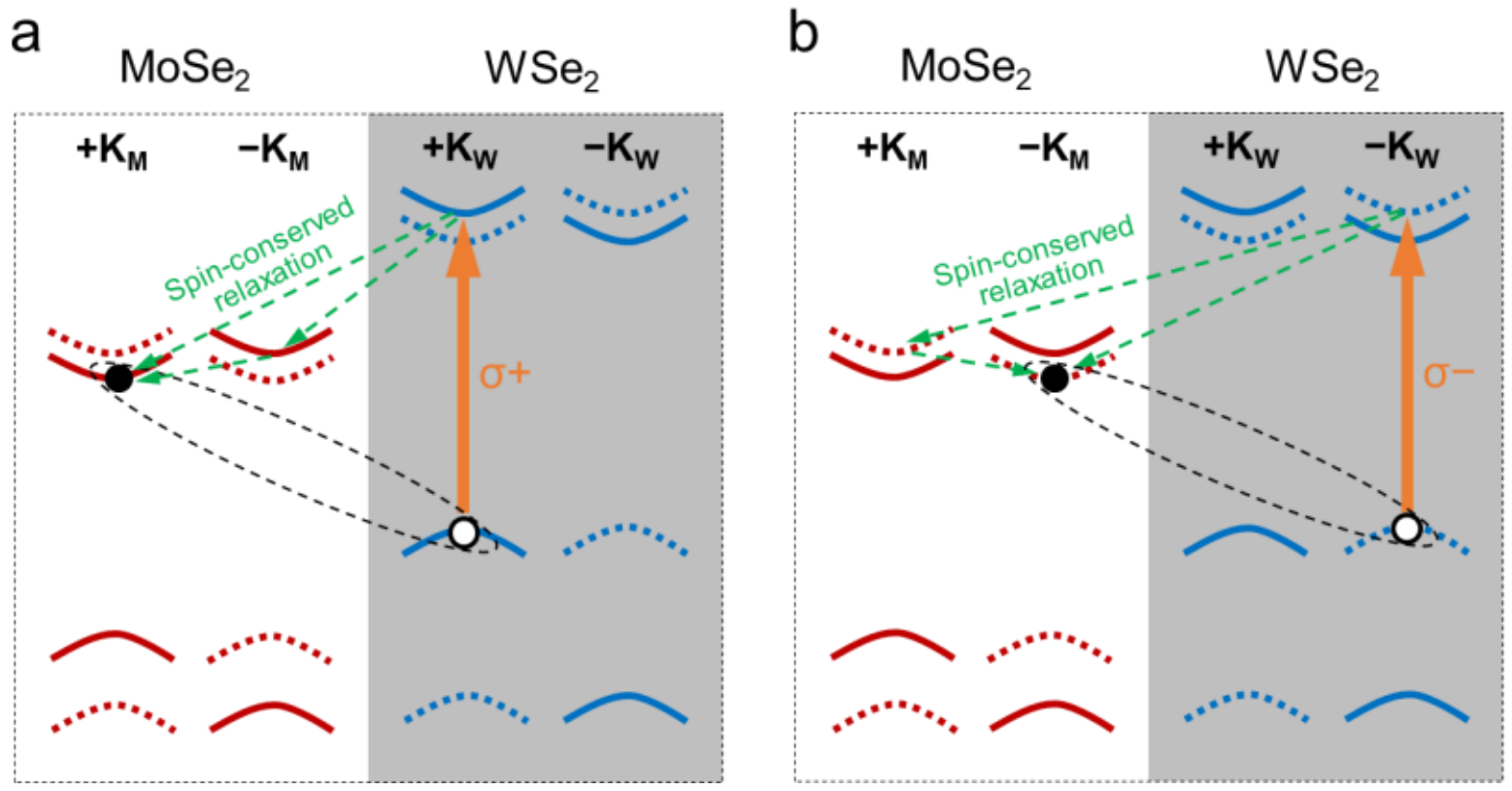

**Supplementary Fig. 7| Spin-conserved relaxation process from intralayer exciton to IX. a,** σ+ and **b,** σ− circularly polarized excitations, which generate $+K_W$ and $-K_W$ intralayer excitons in $WSe_2$, respectively.

## Supplementary Note 2: IX-IX Exchange Interaction and Temperature Dependence

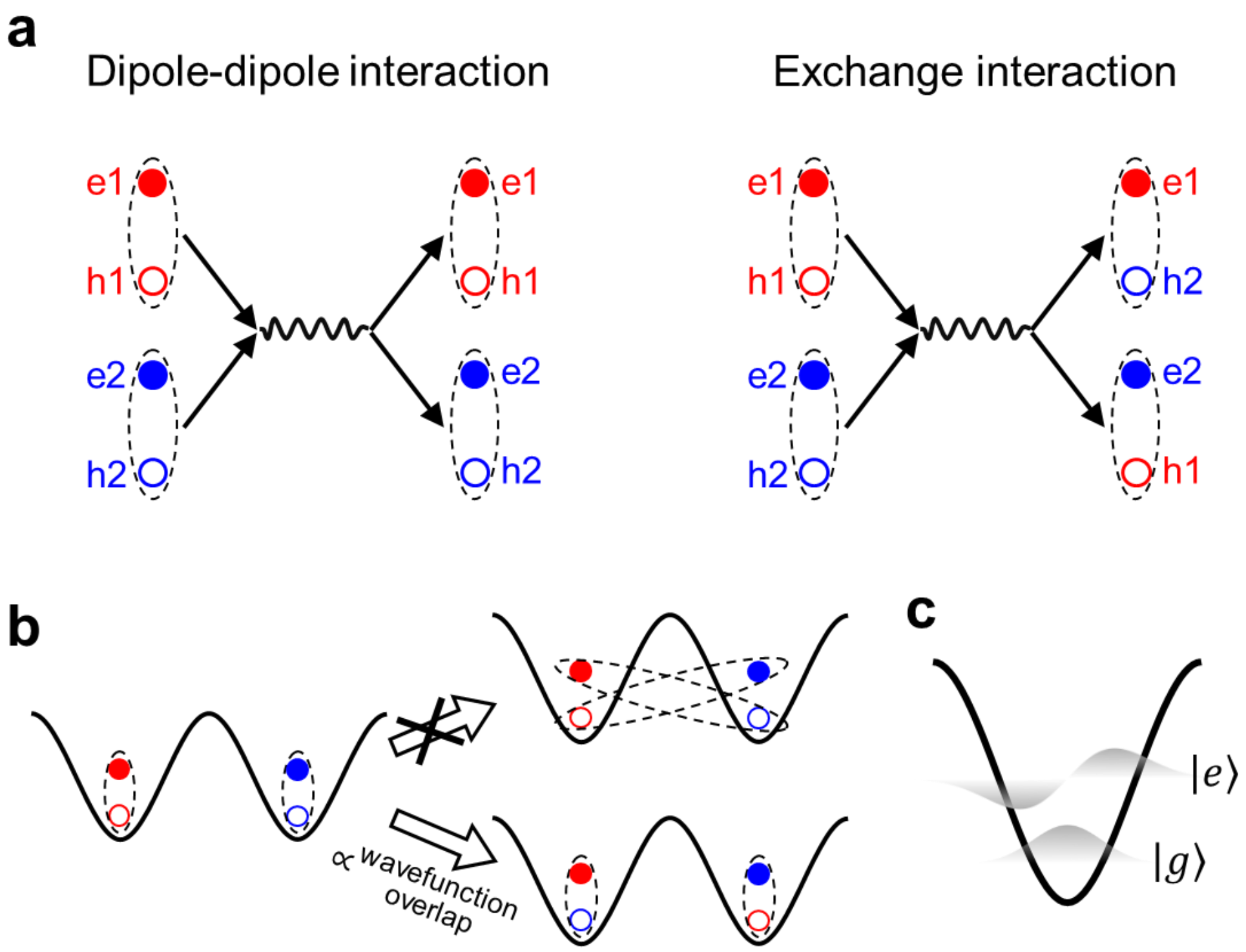

**Supplementary Fig. 8| The IX-IX exchange interaction. a,** Depiction of the dipole-dipole and exchange interactions between two IXs. **b,** The exchange interaction between two IXs trapped at different moiré sites can lead to two consequences. First, IXs with the electron and hole in different sites can form, which, however, is suppressed by their reduced exciton binding energy. Second, due to the wavefunction overlap between the two initial IXs, the exchange interaction can also lead to the formation of IXs with the electron and hole in the same site. **c,** In a moiré trapping potential, the ground state IX $|g\rangle$ has a smaller wavefunction extension than the excited state $|e\rangle$. So the exchange interaction is more significant for the excited states than the ground state.

A simple harmonic trapping model can give a temperature-dependence shown in Supplementary Fig. 9, which is similar to the intensity-temperature curve in Fig 1d. Here we assume the excited states $|e_n\rangle$ are delocalized so they have large nonradiative decay rates (i.e., scattering out of the collection spot). The PL intensity is then mainly from the localized ground state $|g\rangle$, which under equilibrium is proportional to

$$\left[1+\exp\left(-\frac{\Delta}{k_B T}\right)+\exp\left(-\frac{2\Delta}{k_B T}\right)+\cdots\right]^{-1}=1-\exp\left(-\frac{\Delta}{k_B T}\right).$$

The ground state IX intensity as a function of temperature can then be described by a simple equation $\rho_g(T)=\left[1-\exp\left(-\frac{\Delta}{k_B T}\right)\right]\rho_0$, see Supplementary Fig. 9.

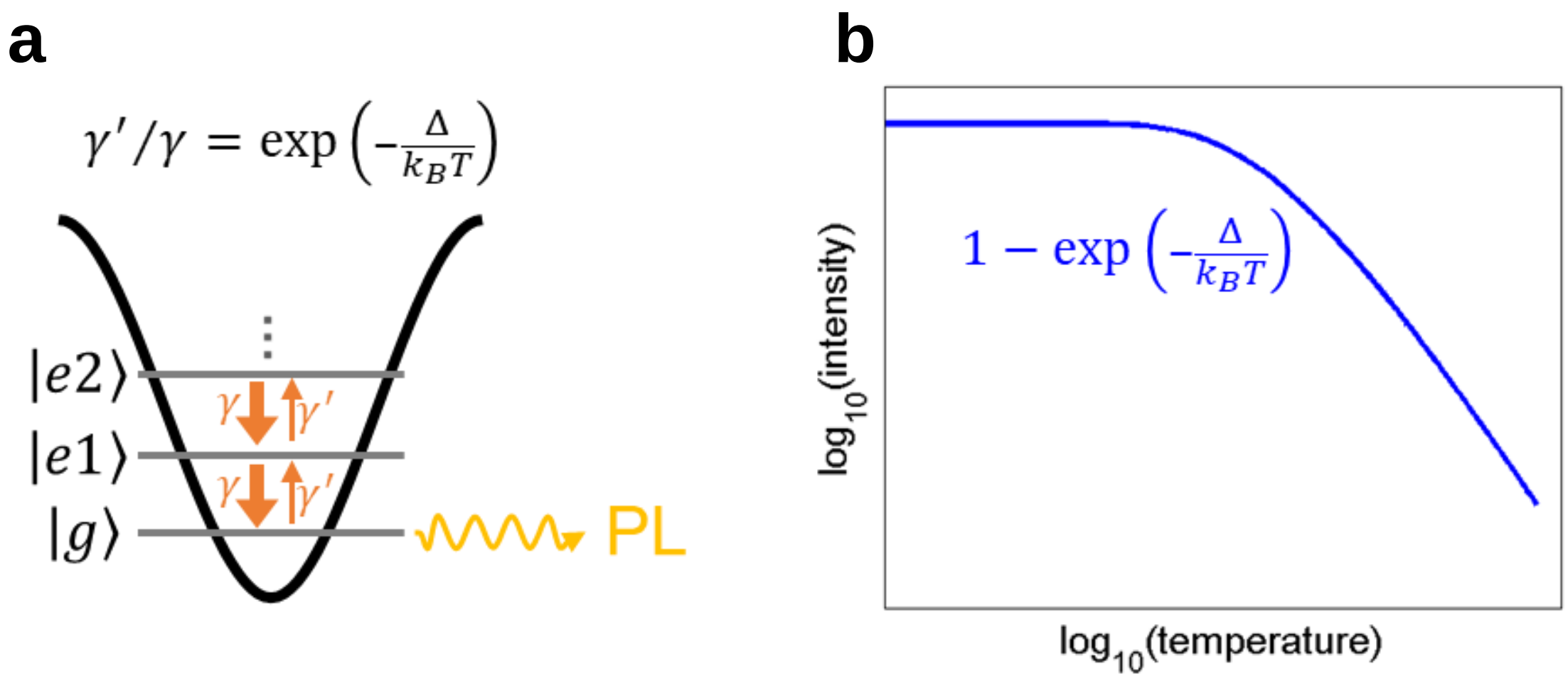

**Supplementary Fig. 9| a,** Depiction of ground and excited states of a harmonic trap. $\Delta$ is the effective level spacing $\gamma$ and $\gamma'$ are the scattering rates. **b,** Ground state IX density as a function of temperature.

Meanwhile, only those ground state IXs in the intravalley form can radiatively recombine, while the other intervalley IXs are dark. The dark intervalley IXs can be scattered into the bright intravalley IXs through the IX-IX exchange interaction, with a temperature and power dependent scattering rate. The fraction of the bright IX in the ground state can be modeled by a temperature and power dependent coefficient, which gives the PL intensity:

$$\rho_{PL}(P,T)=\alpha(P,T)\left[1-\exp\left(-\frac{\Delta}{k_B T}\right)\right]\rho_0.$$

For a phenomenological model, we can write

$$\alpha(P,T) = \alpha(P)\left(1 + \beta(P)\exp\left(-\frac{\Delta}{k_B T}\right)\right),$$

Here $\alpha(P)$ is the power-dependent PL intensity when $T \rightarrow 0$. $\exp\left(-\frac{\Delta}{k_B T}\right)$ comes from the fact that the IX-IX exchange scattering is only significant for the excited states $|e_n\rangle$, but not for the ground state $|g\rangle$. $|\beta(P)|$ accounts for the IX-IX exchange scattering rate which increases with the power. $\beta(P) > 0$ for the H-type sample (the scattering rate from dark to bright IXs increases with temperature), and $\beta(P) < 0$ for the R-type (the scattering rate from bright to dark IXs also increases with temperature).

The normalized PL intensity is then

$$\rho_{\text{normalized}}(P,T) = \left[1 + \beta(P)\exp\left(-\frac{\Delta}{k_B T}\right)\right]\left[1 - \exp\left(-\frac{\Delta}{k_B T}\right)\right].$$

Below is the $\rho_{\text{normalized}}$ curves for four different $\beta(P)$ values. The left with $\beta(P) < 0$ is for R-type, and the right with $\beta(P) > 0$ is for H-type. The single arrows indicate the transition temperature. So this model can qualitatively reproduce the different power dependences of the transition temperature for R- and H-type samples (see Supplementary Fig. 10).

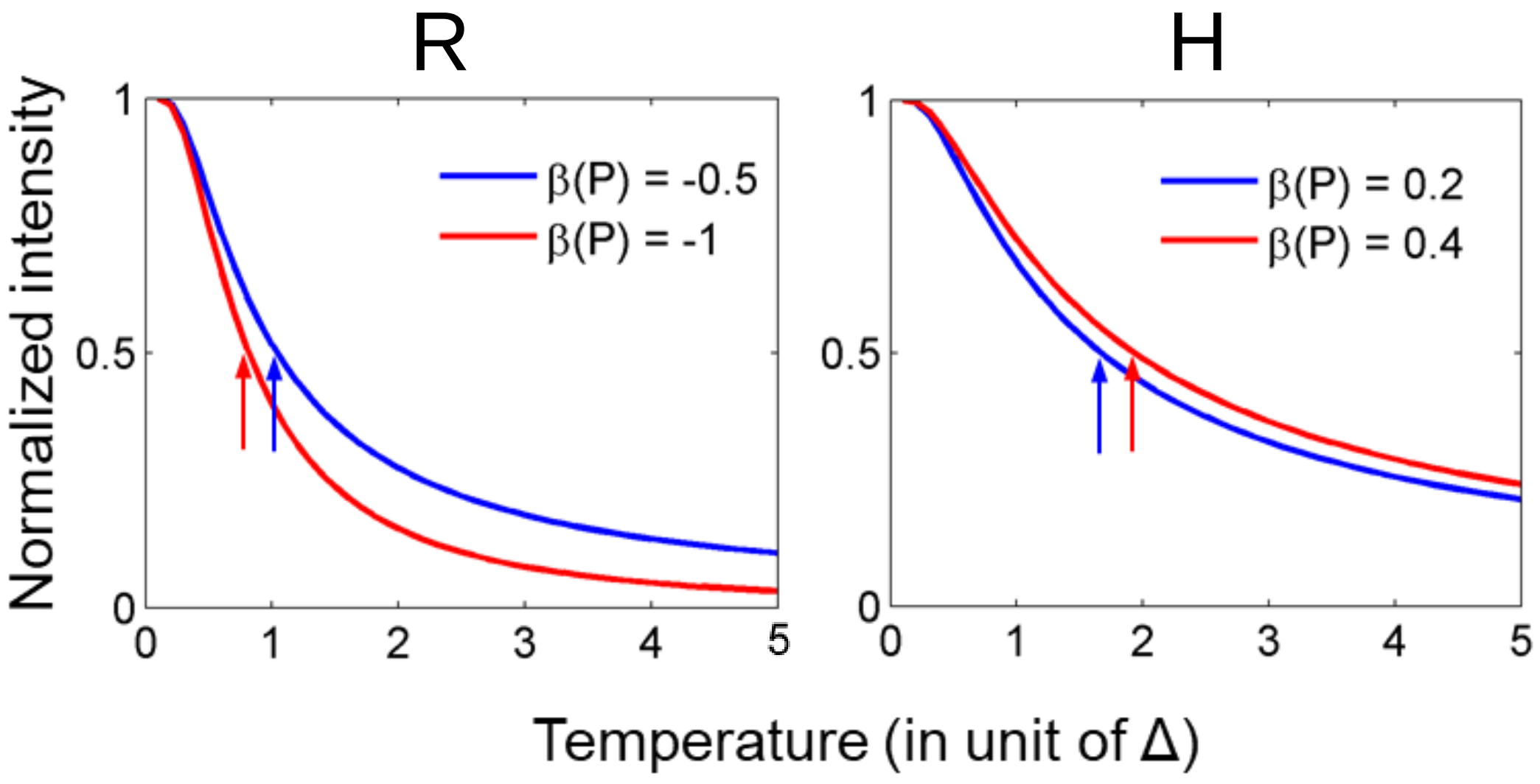

**Supplementary Fig. 10| a,** Theoretical IX PL intensity for two different values of $\beta$ for R-type structure showing the decreasing transition temperature with increasing power. **b,** Theoretical IX PL intensity for two different values of $\beta$ for H-type structure showing the increasing transition temperature with increasing power.